\documentclass[preprint,showpacs,preprintnumbers,amsmath,amssymb]{revtex4}
\usepackage{amsmath}
\usepackage{graphicx}
\usepackage{dcolumn}
\usepackage{bm}

\begin{document}
\title{Generalized Miller Formul\ae}

\author{W. Ettoumi}
\author{Y. Petit}
\author{J. Kasparian}\email{jerome.kasparian@unige.ch}
\author{J.-P. Wolf}
\affiliation{Universit\'e de Gen\`eve, GAP-Biophotonics, 20 rue de l'\'Ecole de M\'edecine, 1211 Geneva 4, Switzerland}

\begin{abstract}
We derive the spectral dependence of the non-linear susceptibility of any order, generalizing the common form of Sellmeier equations. This dependence is fully defined by the knowledge of the linear dispersion of the medium. This finding generalizes the Miller formula to any order of non-linearity. In the frequency-degenerate case, it yields the spectral dependence of non-linear refractive indices of arbitrary order.
\end{abstract}
\pacs{120.4530 Optical constants; 190.0190 Nonlinear optics; 190.3270 Kerr effect; 190.7110 Ultrafast nonlinear optics}


\maketitle

\section{Introduction}
Non-linear optics \cite{Boyd} relies on the knowledge of the non-linear susceptibilities (or, alternatively, the non-linear indices) of the propagation media. This description is generally truncated to the first term, \emph{i.e.} the second-order susceptibility in non-centrosymmetric media, or the third-order susceptibility in centrosymmetric ones.
However, the increase of the available laser powers as well as the investigation of systems like optical fibers \cite{DudleyGC06} or photonic crystals \cite{Soljacic04} where the confinement of the light increases its intensity raise the need to consider higher-order processes. Recently, the non-linear refractive indices of O$_2$, N$_2$ were measured up to $n_8$ (\emph{i.e.} the $9^{th}$-order susceptibility $\chi^{(9)}$) and in Ar up to $n_{10}$ (\emph{i.e.} $\chi^{(11)}$) at 800 nm \cite{LoriotHFL09}. Furthermore, we demonstrated \cite{LoriotBHFLHKW09} that they must be considered in the description of the filamentation of ultrashort pulses \cite{BraunKLDSM95,ChinHLLTABKKS05,BergeSNKW07,CouaironM07,KasparianW08}. This result, which was unexpected, provides a clear illustration of this need, and of the associated requirement to evaluate these terms at any wavelength. But systematic measurements over the spectrum are out of reach of present experimental capabilities. A theoretical support is therefore required to extend the existing experimental results to any frequency, which would provide a new insight into non-linear optics.

Such relation was provided by Miller in the case of the second-order susceptibility. He observed that in many crystals the knowledge of  $\chi^{(2)}$ for one single triplet of frequencies $(\omega'_0;\omega'_1,\omega'_2)$ and the dispersion relation of the medium is sufficient to determine  $\chi^{(2)}$ for any triplet of frequencies $(\omega_0;\omega_1,\omega_2)$ \cite{Miller64}:
\begin{equation}
\frac{\chi^{(2)}(\omega_0;\omega_1,\omega_2)}{\chi^{(2)}(\omega_0';\omega'_1,\omega'_2)}
= \frac{\chi^{(1)}(\omega_0)\chi^{(1)}(\omega_1)\chi^{(1)}(\omega_2)} {\chi^{(1)}(\omega_0')\chi^{(1)}(\omega'_1)\chi^{(1)}(\omega'_2)}
\label{Miller}
\end{equation}

This formula has been widely used \emph{e.g.} determine the to derive the second order static susceptibility in semiconductors \cite{ScandoloB95}. Mizrahi and Shelton later suggested that the Miller formula could be extended to provide the spectral dependence of the non-linear index $n_2$, \emph{i.e.} to the third-order frequency-degenerate non-linearity \cite{MizrahiS85}. However, they did not demonstrate this result in their article, but rather referred to an unpublished work \cite{Owyoung72}. Later, Bassani \emph{et al.} demonstrated this relation for any order of non-linearity in the specific case of harmonic generation processes \cite{BassaniL98, LucariniBPS03}.

In this Letter, provide an explicit expression of the spectral dependence of the non-linear susceptibility of any order, generalizing the common form of Sellmeier equations. We show that this spectral dependence is fully defined by the knowledge of the linear dispersion of the medium. As a consequence, the Miller formula (\ref{Miller}) \cite{Miller64} can be extended to any order of non-linearity. In the frequency-degenerate case, this finding yields the spectral dependence of non-linear refractive indices of arbitrary order, confirming and widely generalizing Mizrahi and Shelton's statement about $\chi^{(3)}$ \cite{MizrahiS85}.

\section{Derivation of generalized Miller formul\ae}
\subsection{Electrons in an anharmonic oscillator}

Elucidating the spectral dependence of the electric susceptibility requires to consider the equation of motion of an electron located at $\vec{r}$ in a three-dimensional potential $V(\vec{r})$. $V$ can be expanded as a 3-dimensional Taylor series around the equilibrium position $\vec{r}=\vec{0}$:
\begin{equation}
\displaystyle{V(\vec{r}) = V(\vec{0}) + \sum_{i+j+k \ge 2}{\frac{x^i y^j z^k}{i!j!k!} \left[\frac{\mathrm{d}^{i+j+k} V}{\partial x^i \partial y^j \partial z^k}\right]_{\vec{r}=\vec{0}}}}
\end{equation}
where the summation begins at $q\equiv i+j+k=2$ since $\left[\frac{\partial V}{\partial x}\right]_{x=0}=\left[\frac{\partial V}{\partial y}\right]_{y=0}=\left[\frac{\partial V}{\partial z}\right]_{z=0}=0$ by definition of the equilibrium position.
As a consequence, the electron experiences a force equal to
\begin{equation}
\vec{F}(\vec{r}) = -\vec{\nabla} V(\vec{r})= - \sum_{q \ge 2}{\begin{pmatrix}
\frac{x^{i-1} y^j z^k}{(i-1)!j!k!} \\
\frac{x^{i} y^{j-1} z^k}{i!(j-1)!k!}  \\
\frac{x^{i} y^j z^{k-1}}{i!j!(k-1)!}
\end{pmatrix}} \left[\frac{\partial^{q} V}{\partial x^i \partial y^j \partial z^k}\right]_{\vec{r}=\vec{0}}
\label{force}
\end{equation}

The macroscopic polarization along the axes $x$, $y$ and $z$ are respectively $P_x=-Nex$, $P_y=-Ney$ and $P_z=-Nez$, $N$ being the local density of electrons and $-e$ their charge. Equation (\ref{force}) therefore rewrites:
\begin{equation}
\vec{F}(\vec{r}) = -\sum_{q \ge 2}{\left(\frac{-1}{Ne}\right)^{q-1} \begin{pmatrix}
\frac{P_x^{i-1} P_y^j P_z^k}{(i-1)!j!k!} \\
\frac{P_x^{i} P_y^{j-1} P_z^k}{i!(j-1)!k!} \\
\frac{P_x^{i} P_y^j P_z^{k-1}}{i!j!(k-1)!}
\end{pmatrix}} \left[\frac{\partial^{q} V}{\partial x^i \partial y^j \partial z^k}\right]_{\vec{r}=\vec{0}}
\end{equation}
or, equivalently:
\begin{equation}
\vec{F}(\vec{r}) = -\sum_{q \ge 1}\left(\frac{-1}{Ne}\right)^{q} \frac{P_x^{i} P_y^j P_z^k}{i!j!k!} 
\begin{pmatrix}
\left[\frac{\partial^{q+1} V}{\partial x^{i+1} \partial y^j \partial z^k}\right]_{\vec{r}=\vec{0}} \\
\left[\frac{\partial^{q+1} V}{\partial x^i \partial y^{j+1} \partial z^k}\right]_{\vec{r}=\vec{0}} \\
\left[\frac{\partial^{q+1} V}{\partial x^i \partial y^j \partial z^{k+1}}\right]_{\vec{r}=\vec{0}}
\end{pmatrix}
\end{equation}

We introduce the $(q+1)^{th}$ rank tensor $Q^{(q)}$, which elements are given by:
\begin{equation}
\begin{pmatrix} Q^{(q)}_{x;x^{(i)},y^{(j)},z^{(k)}} \\ Q^{(q)}_{y;x^{(i)},y^{(j)},z^{(k)}} \\ Q^{(q)}_{z;x^{(i)},y^{(j)},z^{(k)}} \end{pmatrix}
\equiv \frac{(-1)^{q}}{m \times i!j!k!}
\begin{pmatrix}
\left[\frac{\partial^{q+1} V}{\partial x^{i+1} \partial y^j \partial z^k}\right]_{\vec{r}=\vec{0}} \\
\left[\frac{\partial^{q+1} V}{\partial x^i \partial y^{j+1} \partial z^k}\right]_{\vec{r}=\vec{0}} \\
\left[\frac{\partial^{q+1} V}{\partial x^i \partial y^{j} \partial z^{k+1}}\right]_{\vec{r}=\vec{0}}
\end{pmatrix}
\end{equation}
where, for example, $x^{(i)}$ indicates that the coordinate $x$ appears $i$ times in the index list of the considered tensor element. As a consequence, the classical equation of motion of the electron becomes:
\begin{equation}
\frac{\mathrm{d}^2 \vec{P}}{\mathrm{d}t^2} +
\bar{\bar{\gamma}} \frac{\mathrm{d} \vec{P}}{\mathrm{d}t} + \bar{\bar{\omega}}_{e}^2 \vec{P}
+ N e \sum^{\infty}_{q=1} {Q^{(q)} : \bigotimes_{l=1}^{q} \frac{\vec{P}}{Ne}}=\frac{N e^2}{m} \vec{E}(t)
\label{eqn:P_tensoriel}
\end{equation}
where $:$ denotes the contracted product and $\bigotimes$ the tensorial product.
$\vec{E}$ is the driving electric field, $\bar{\bar{\omega}}_{e}^2 = \begin{pmatrix}
\omega_{e,x}^2&0&0\\
0&\omega_{e,y}^2&0\\
0&0&\omega_{e,z}^2
\end{pmatrix}$
is the eigenfrequency matrix of the considered medium, which is diagonal provided $x$, $y$, and $z$ are the principal axes of the optical frame of the medium (\emph{e.g.} $\omega_{e,x}=\sqrt{\frac{1}{m}\frac{\partial^2V}{\partial x^2}}$), and the matrix $\bar{\bar{\gamma}}$ stands for the linear absorption. No multiphoton absorption is considered here. Note that a medium symmetry of at least K$\times$C$_2$ (\emph{i.e.} a crystal with at least orthorhombic symmetry, or a statistically isotropic medium like air) allows to simultaneously diagonalize $\bar{\bar{\omega}}_{e}^2$ and $\bar{\bar{\gamma}}$ in the optical frame \cite{BoulangerPSFMZA08}.

A perturbative solution of equation (\ref{eqn:P_tensoriel}) is searched as:
\begin{equation}
\displaystyle{\vec{P} = \sum^{\infty}_{l=1} \alpha^{l} \vec{P}^{(l)}}
\end{equation}
where $\alpha \in ]0..1[$ is a free parameter. The series begins at 1, assuming $\vec{P}^{(0)}=\vec{0}$. Inserting it in equation (\ref{eqn:P_tensoriel}) and equating the terms in $\alpha^{q}$ yields a set of equations
\begin{equation}
\frac{\mathrm{d}^2 \vec{P}^{(1)}}{\mathrm{d}t^2} + \bar{\bar{\gamma}} \frac{\mathrm{d} \vec{P}^{(1)}}{\mathrm{d} t} +
\bar{\bar{\omega}}_{e}^2 \vec{P}^{(1)} = \frac{N e^2}{m} \vec{E}(t) \label{eqn:ordre1_vectoriel}\\
\end{equation}
and, $\forall q \ge 2$,
\begin{equation}
\frac{\mathrm{d}^2 \vec{P}^{(q)}}{\mathrm{d}t^2} + \bar{\bar{\gamma}} \frac{\mathrm{d} \vec{P}^{(q)}}{\mathrm{d} t} +
\bar{\bar{\omega}}_{e}^2  \vec{P}^{(q)}
= - N e Q^{(q)}: \bigotimes_{u=1}^{q} \frac{\vec{P}^{(1)}}{Ne} \label{eqn:systeme_tensoriel}
\end{equation}

Note that in the right-hand side of Equation (\ref{eqn:systeme_tensoriel}), we have deliberately omitted the terms in $\bigotimes_{\sum{q'_u}=q}\left(P^{(q'_u)}\right)$, with $q>q'_u>1$. These terms imply non-linear polarizations in the construction of the considered higher-order non-linear polarization and hence correspond to cascades of frequency mixings, like \emph{e.g.} in the generation of third harmonic by frequency-doubling the fundamental wavelength and then mixing it with its second-harmonic. Consistently, in the identification of the non-linear susceptibilities in Equation (\ref{P_chik}), only single-step mixing will be taken into account.

\subsection{Sellmeier equations}
If $\bar{\bar{\Omega}}$ can be diagonalized (\emph{i.e.} if the medium symmetry is at least K$\times$C$_2$ or absorption can be neglected), Equation (\ref{eqn:ordre1_vectoriel}) defining the linear polarization is purely vectorial and can be solved on each axis independently. Omitting the indices displaying the axis of the considered polarization, we write $P^{(1)}(t)=\int_{-\infty}^{+\infty}{P^{(1)}_0(\omega) e^{i \omega t}d\omega}$ and $E(t)=\int_{-\infty}^{+\infty}{E_0({\omega}) e^{i{\omega} t} d{\omega}}$, where, following the usual notation, positive frequencies denote incident ones and negative frequencies are emitted ones. Introducing the spectral dependency parameter $\Omega(\omega) = {\omega_e}^2-\omega^2+i\omega\gamma$ yields
\begin{equation}
\int_{-\infty}^{+\infty}{\Omega(\omega)P_0^{(1)}(\omega)e^{i\omega t}d\omega}=\frac{N e^2}{m}\int_{-\infty}^{+\infty}{E_0({\omega}) e^{i{\omega} t} d{\omega}}
\label{OmegaP_E}
\end{equation}
which implies, for any frequency $\omega_0$ of the emitted field except for the resonance frequency  :
\begin{equation}
P^{(1)}_{0}(\omega_0) = \frac{N e^2}{m} \: \frac{E_0(\omega_0)}{\Omega(\omega_0)}
\end{equation}

Hence, the linear susceptibility at frequency $\omega_0$ is:
\begin{equation}
\chi^{(1)}(\omega_0) = \frac{1}{\Omega(\omega_0)} \frac{N e^2}{m \epsilon_0} = \frac{N e^2}{m \epsilon_0({\omega_e}^2-\omega_0^2+i\omega_0\gamma)}  \label{Sellmeier1}
\end{equation}
which defines the spectral dependence of $\chi^{(1)}$ on any principal axis. If absorption can be neglected ($\gamma\sim0 $) and $\chi^{(1)} \ll 1$, Equation (\ref{Sellmeier1}) takes a form similar to a typical Sellmeier formula:  $n^2-1=\frac{A}{B-1/ \lambda^2}$, where $n^2=1+\chi^{(1)}$.

\subsection{Uniaxial generalized Miller formul\ae}
Equation (\ref{eqn:systeme_tensoriel}) can be reduced to a scalar form provided the polarization is excited along one single principal axis. We will first focus on this case which simplifies the writing and thus helps focusing the discussion on the the physical aspects, without altering the principle of the derivation. Omitting the index corresponding to the axis, the scalar form of Equation (\ref{eqn:systeme_tensoriel}) reads, for any $q \ge 2$:
\begin{equation}
\frac{\mathrm{d}^2 P^{(q)}}{\mathrm{d}t^2} + \gamma \frac{\mathrm{d} P^{(q)}}{\mathrm{d} t} + {\omega_e}^2 P^{(q)} = -NeQ^{(q)} \left(\frac{P^{(1)}}{N e}\right)^q
\label{eqn:systeme}
\end{equation}

Since $P^{(q)}(t)=\int_{-\infty}^{+\infty}{P^{(q)}_0(\omega) e^{i \omega t}d\omega}$, Equation (\ref{eqn:systeme}) rewrites:
\begin{eqnarray}
\int_{-\infty}^{+\infty}\Omega(\omega)P_0^{(q)}(\omega)e^{i\omega t}d\omega 
&=& -\frac{Q^{(q)}}{(Ne)^{q-1}} \left(\int_{-\infty}^{+\infty}{P_0^{(1)}(\omega)e^{i\omega t}d\omega}\right)^q \\
&=& -\frac{Q^{(q)}}{(Ne)^{q-1}} \iint..\iint \prod_{l=1}^q \left(P_0^{(1)} (\omega_l) e^{i\omega_l t}d\omega_l\right)
\end{eqnarray}

Identifying the terms at an arbitrary frequency $\omega_0$ on both sides of the equation, we obtain:
\begin{eqnarray}
P^{(q)}_{0}(\omega_0) = -\frac{Q^{(q)}}{\Omega(\omega_0)(Ne)^{q-1}} \iint..\iint \delta\left(\sum_{l=0}^q{\omega_l}=0\right) 
\times \prod_{l=1}^q \left({P^{(1)}(\omega_l)d\omega_l}\right)
\end{eqnarray}
\begin{equation}
P^{(q)}_{0}(\omega_0) = -Ne\left(\frac{e}{m}\right)^{q}\frac{Q^{(q)}}{\Omega(\omega_0)} \sum_{\sum_{l=0}^q{\omega_l}=0} \left(\prod_{l=1}^q{\frac{E_0(\omega_l)}{\Omega(\omega_l)}}\right)
\label{eqn:polar}
\end{equation}

Let us now consider the construction of a wave at $\omega_0$ from a set of $q' \leq q$ incident waves at frequencies $\omega_1$,...,$\omega_{q'}$, each algebric frequency $\omega_l$ being implied  $u_l$ times in the process: $\sum_{l=1}^{q'}{u_l}=q$, while energy conservation imposes $\sum_{l=1}^{q'}{u_l\omega_l}=\omega_0$. In this case,
\begin{equation}
P^{(q)}_{0}(\omega_0) = - Ne\left(\frac{e}{m}\right)^{q}\frac{Q^{(q)}}{\Omega(\omega_0)} C_q^{u_1,..,u_{q'}} \prod_{l=1}^{q'}{\left(\frac{E_0(\omega_l)}{\Omega(\omega_l)}\right)^{u_l}}
\end{equation}
where $C_q^{u_1,..,u_{q'}}={q!}/{\left(u_1!\times..\times u_{q'}!\right)}$ is the number of combinations achievable from $q'$ sets of $u_1,..,u_{q'}$ objects, respectively. The terms of Equation (\ref{eqn:polar}) corresponding to each combination of frequencies conserving energy can be identified with the expression of the non-linear polarization using the $(q+1)^{th}$ rank tensor.
\begin{equation}
P^{(q)}_{0}(\omega_0) = \epsilon_0 C_q^{u_1,..,u_{q'}} \chi^{(q)}(\omega_0;\omega_1,...,\omega_q) \prod_{l=1}^{q'}{E_0^{u_l}(\omega_l)}
\label{P_chik}
\end{equation}

As stated above, consistently with Equation (\ref{eqn:systeme}) we do not consider here the cascaded processes. Therefore, the identification yields:
\begin{eqnarray}
\chi^{(q)}(\omega_0;\omega_1,...,\omega_q)&=&-Ne\left(\frac{e}{m}\right)^{q}\frac{Q^{(q)}}{\epsilon_0 \prod_{l=0}^q{\Omega^(\omega_l)}}  \label{chi_k_general} \\
&=&\frac{m\epsilon_0^q}{N^q e^{q+1}}Q^{(q)}\prod_{l=0}^q{\chi^{(1)}(\omega_l)}  \label{chi_k_general_chi1}
\end{eqnarray}

This expression provides a general description of the non-linear susceptibility of any order, provided the shape of the potential $V$ is known. In practice, this potential is rarely known, but it is independent from the excitation frequency. The spectral dependence of the above expression is therefore only driven by $\Omega$, \emph{i.e.} the spectral dependence of the first-order susceptibilities. As a consequence, in a given medium, \emph{the knowledge of both the frequency dependence of the linear susceptibility and of the $q^{th}$-order susceptibility for a specific set of wavelengths is sufficient to extrapolate this susceptibility to any other set of wavelengths}, through the relation:
\begin{equation}
\frac{\chi^{(q)}(\omega_0;\omega_1,...,\omega_q)}{\chi^{(q)}(\omega_0';\omega'_1,...,\omega'_q)}
= \frac{\prod_{l=0}^q{\Omega(\omega'_l)}}{\prod_{l=0}^q{\Omega(\omega_l)}}
= \frac{\prod_{l=0}^q{\chi^{(1)}(\omega_l)}}{\prod_{l=0}^q{\chi^{(1)}(\omega'_l)}}  \label{Miller_general}
\end{equation}

When applied to $\chi^{(2)}$, this generalized Miller formula immediately reduces to the original one of Equation (\ref{Miller}) \cite{Miller64}.

\subsection{Three-dimensionnal Miller formul\ae}
While the one-dimensional treatment of Equations (\ref{eqn:systeme})-(\ref{Miller_general}) provides an easy writing of the derivation, the same sequence can be applied to solve Equation (\ref{eqn:systeme_tensoriel}) in the general three-dimensional case. Products simply have to be transposed to the adequate tensorial products. The identification of the spectral components of the Fourier expression of $\vec{P}^{(q)}$ and $\bigotimes \vec{P}^{(1)}$ yields a three-dimensional equivalent of Equation (\ref{eqn:polar}):
\begin{equation}
\bar{\bar{\Omega}}(\omega_0) \vec{P}^{(q)}_{0}(\omega_0) = - Ne \left(\frac{e}{m}\right)^{q}  {{Q}^{(q)} : \left( \bigotimes_{l=1}^{q} \frac{\vec{P}^{(1)}(\omega_l)}{Ne}\right)}
\end{equation}


In media with at least K$\times$C$_2$ symmetry, or if absorption can be neglected, $\bar{\bar{\Omega}}$ is diagonal so that this expression can easily be identified with the counterpart of Equation (\ref{P_chik}) expressing the non-linear polarization in terms of non-linear susceptibility:
\begin{equation}
\vec{P}^{(q)}_{0}(\omega_0) = \epsilon_0 \chi^{(q)} : \bigotimes_{l=1}^{q} \vec{E}(\omega_l)
\end{equation}
\begin{eqnarray}
\frac{\chi^{(q)}_{v;x^{(i)},y^{(j)},z^{(k)}}(\omega_0;\omega_1,...,\omega_{q})}{\chi^{(q)}_{v;x^{(i)},y^{(j)},z^{(k)}}(\omega_0';\omega'_1,...,\omega'_{q})}
&=& \frac{\prod_{l=0}^{q}{\Omega_{v_l}(\omega'_l)}}{\prod_{l=0}^{q}{\Omega_{v_l}(\omega_l)}}  \\
&=& \frac{\prod_{l=0}^{q}{\chi_{v_l}^{(1)}(\omega_l)}}{\prod_{l=0}^{q}{\chi_{v_l}^{(1)}(\omega'_l)}} \label{Miller_general_tenseur}
\end{eqnarray}
where each index $v_l$ denotes either the $x$, $y$, or $z$ axis. The spectral dependence of $\chi^{(q)}$ only depends on the product of the $\Omega_{v_l}(\omega_l)$, meaning that all frequencies commute. This property, together with the fact that the $\Omega(\omega)$ are even functions as soon as absorption can be neglected, provides a direct evidence and generalization to arbitrary orders of the ABDP relations. These relations state that as soon as $\bar{\bar{\Omega}}$ can be diagonalized, $\omega_0$ commutes with any $\omega_l$ in the expression of $\chi^{(2)}$ and $\chi^{(3)}$ \cite{Armstrong62}. 

\subsection{Application to non-linear refractive indices}
\begin{figure}[ht]
  \begin{center}
      \includegraphics[keepaspectratio, width=6cm]{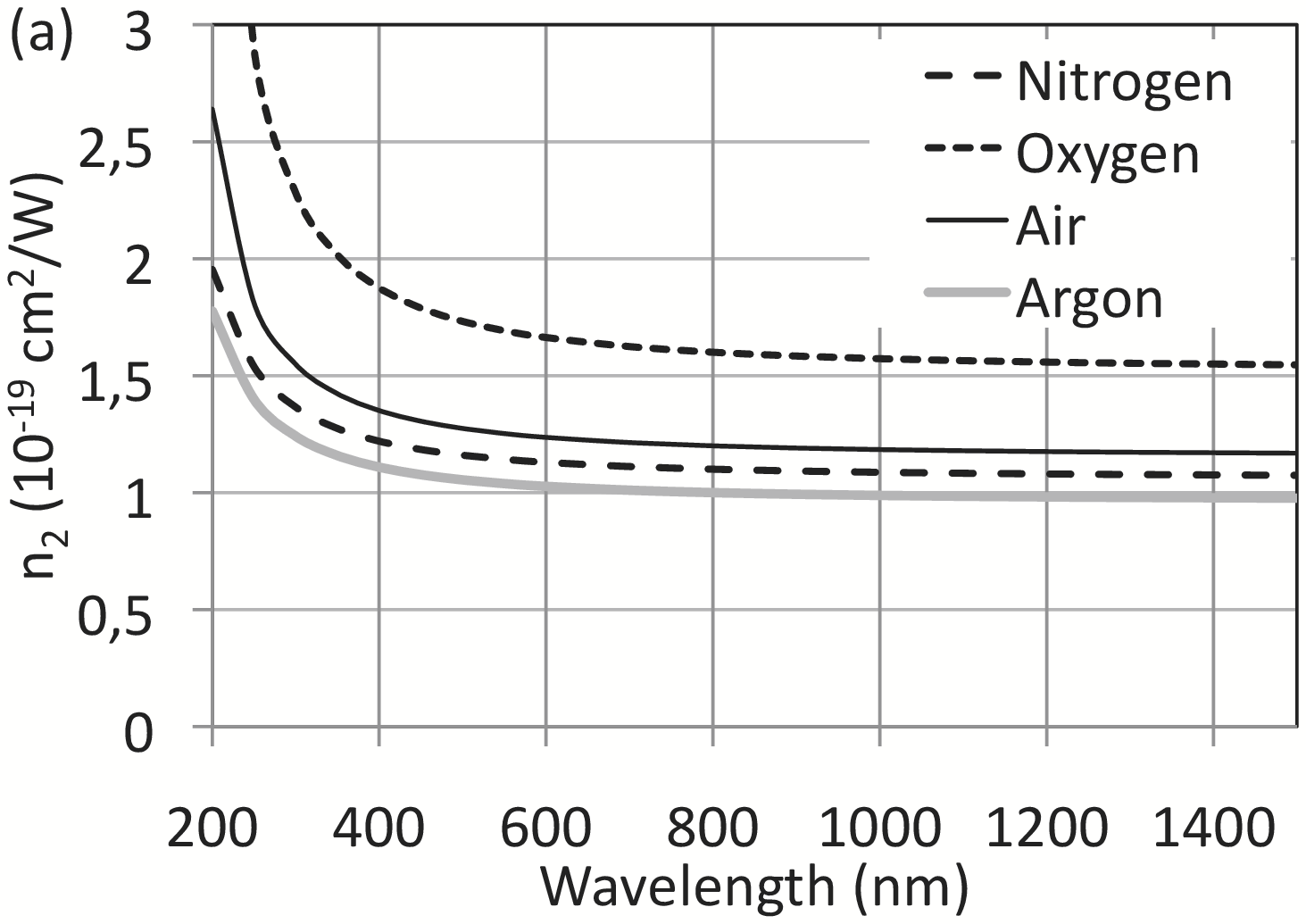}
      \includegraphics[keepaspectratio, width=6cm]{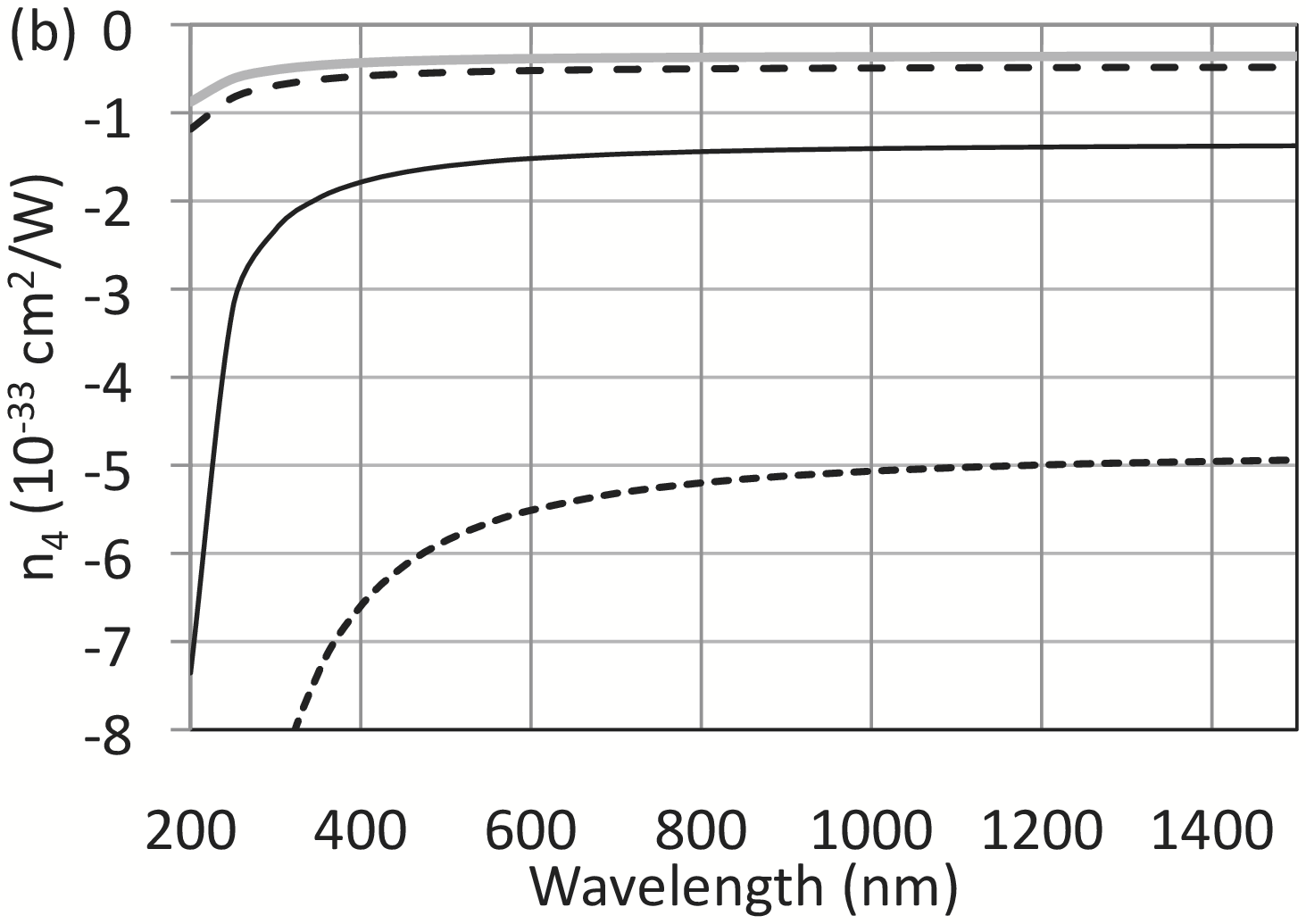}
      \includegraphics[keepaspectratio, width=6cm]{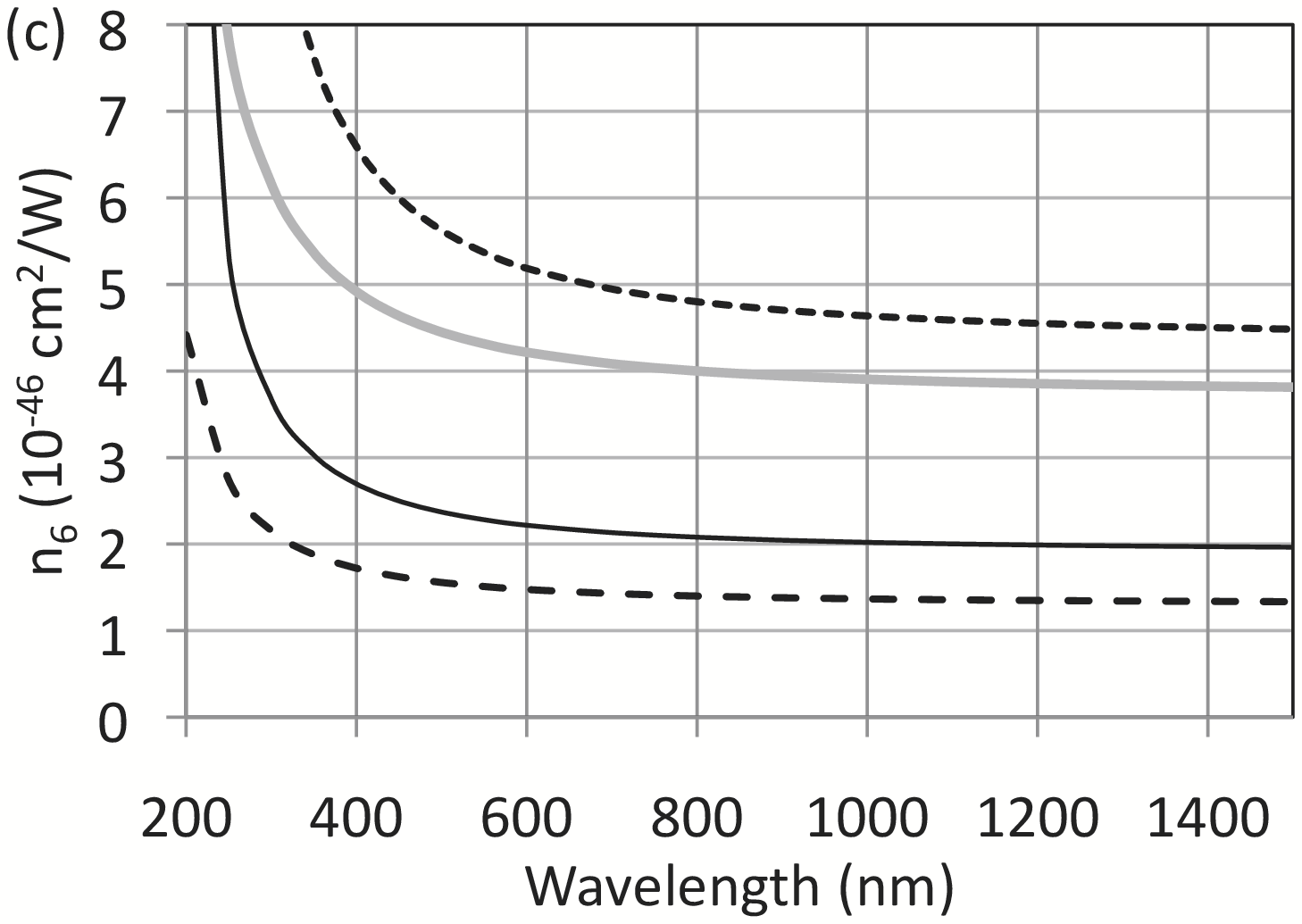}
      \includegraphics[keepaspectratio, width=6cm]{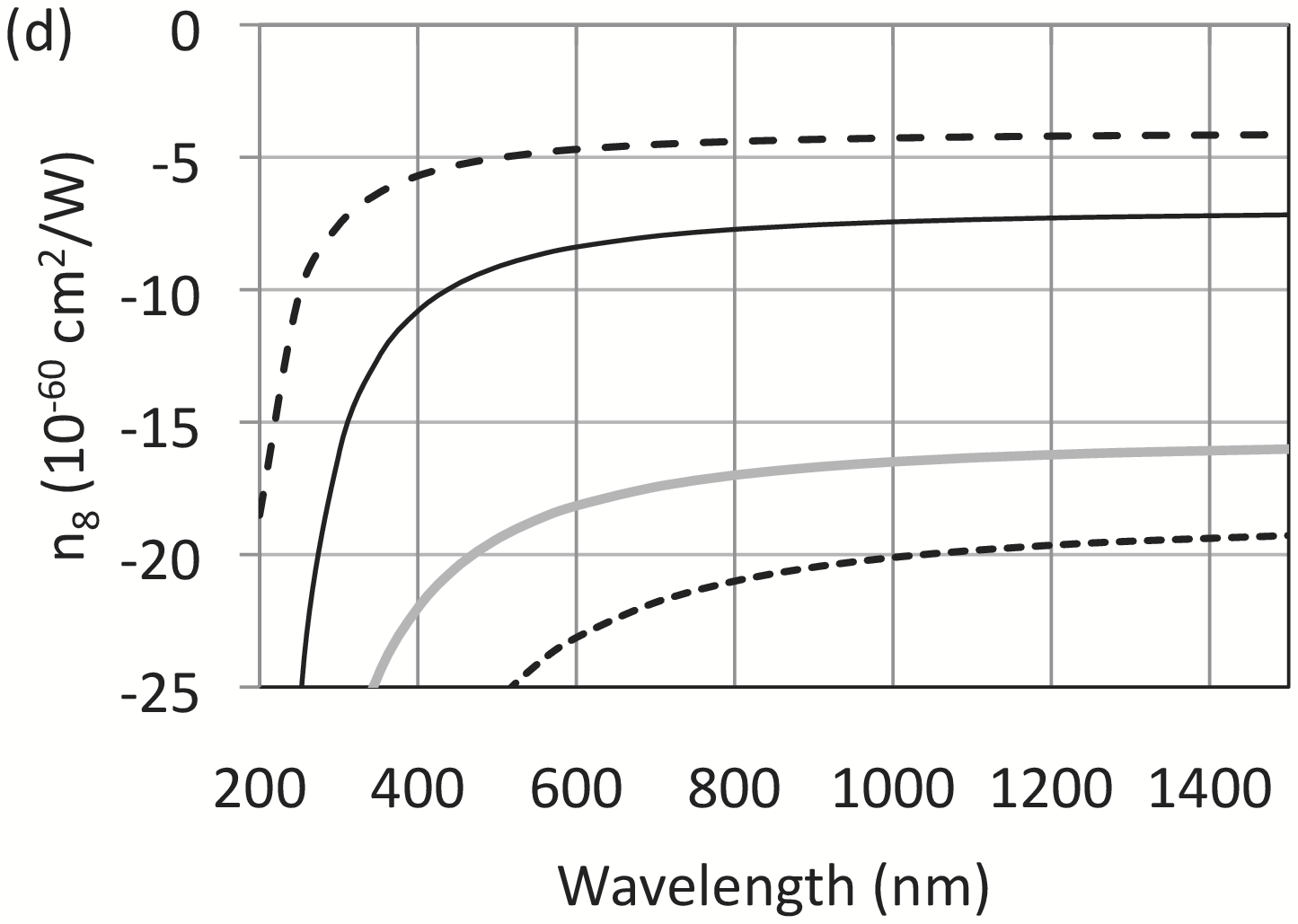}
  \end{center}
  \caption{Spectral dependence of the non-linear refractive indices (a) $n_2$, (b) $n_4$, (c) $n_6$, and (d) $n_8$ of O$_2$, N$_2$, air and Ar at atmospheric pressure}
  \label{n_i_spectral}
\end{figure}

In the case of frequency-degenerate interactions of odd order excited along one single principal axis, the susceptibility tensor reduces to $\chi^{(2p+1)}(\omega_0) \equiv \frac{2^{p+1}p!(p+1)!}{(2p+1)!}\frac{\left(n_0^2(\omega_0)\epsilon_0 c^2\right)^p}{n_0(\omega_0)}n_{2p}(\omega_0)$, which defines the $p^{th}$-order non-linear refractive index $n_{2p}$. Therefore, the knowledge of the dispersion curve for the linear susceptiblity and the measurement of $\chi^{(2p+1)}$ at one single frequency $\omega'_0$ are sufficient to provide $\chi^{(2p+1)}$ at any frequency $\omega_0$ thanks to Equation (\ref{Miller_general}):
\begin{equation}
\frac{\chi^{(2p+1)}(\omega_0)}{\chi^{(2p+1)}(\omega_0')}
= \left(\frac{\Omega(\omega_0')}{\Omega(\omega_0)}\right)^{2p+2}
= \left(\frac{\chi^{(1)}(\omega_0)}{\chi^{(1)}(\omega_0')}\right)^{2p+2}  \label{Miller_general_degenere}
\end{equation}
or equivalently, if absorption can be neglected, in terms of the real parts of the non-linear refractive indices:
\begin{equation}
\frac{n_{2p}(\omega_0)}{n_{2p}(\omega_0')}= \left(\frac{n^{2}_{0}(\omega_0)-1}{n^{2}_{0}(\omega_0')-1}\right)^{2p+2}
\label{miller_n_k}
\end{equation}
As an illustration, Figure \ref{n_i_spectral} displays the values of $n_2$ through $n_8$ for N$_2$, O$_2$ and Ar, based on the recent experimental measurements of Loriot \emph{et al.} at 800 nm \cite{LoriotHFL09,LoriotBHFLHKW09}, extrapolated to the whole spectrum using Equation (\ref{miller_n_k}) and the dispersion data of Zhang \emph{et al.} \cite{ZhangLW08}. These dispersion data have been chosen because they have been measured consistently for all proposed species. However, we checked that they agree within 2\% with typical data available over the spectral range plotted \cite{PeckK66,PeckR72,Birch94}.

\section{Generalization to mixes and multiple resonance frequencies}
Up to now, we have only considered the case of a single type of oscillators. However, in actual media (\emph{e.g.} in mixes like the air, where both N$_2$ and O$_2$ get polarized, or in crystals where several oscillation modes can be excited) several eigenfrequencies $\omega_{e,l}$ may contribute.  In this case, the polarization $\vec{P}$ can be defined as $\sum{\vec{P}_l}$, where $l$ denotes the types of oscillators, or species. If the coupling between the oscillators can be neglected, the above derivation applies to each oscillator type independently, and the resulting refractive index will be given by the Lorentz-Lorenz model. For example, away from the resonances, the refractive index of air is given by the Sellmeier equation \cite{ZhangLW08}:
\begin{equation}
10^8 (n_0 - 1) = 8015.514 + \frac{2368616}{128.7459 - 1/\lambda^2} + \frac{19085.73}{50.01974 - 1/\lambda^2}
\label{air}
\end{equation}
where the wavelength $\lambda$ is expressed in $\mathrm{\mu}$m. Obviously, the terms of Equation (\ref{air}) respectively correspond to N$_2$, with a resonance at $1/\lambda^2 \sim 128.7 \mu$m$^{-2}$ (\emph{i.e.} $\lambda = 88$~nm):
\begin{equation}
10^8 (n_{0,N_2} - 1) = 8736.28 + \frac{2398095.2}{128.7 - 1/\lambda^2} \label{N2}\\
\end{equation}
and to O$_2$, with $1/\lambda^2 \sim 50 \mu$m$^{-2}$ (\emph{i.e.} $\lambda = 141$~nm) \cite{ZhangLW08}:
\begin{equation}
10^8 (n_{0,O_2} - 1) = 15532.45 + \frac{456402.97}{50.0 - 1/\lambda^2} \label{O2}
\end{equation}

In such cases, the generalized Miller formul\ae\ of Equation (\ref{Miller_general}) cannot be applied to the material or the mix as a whole. Instead, they must be applied to each susceptibility-order of each oscillator type. The non-linear susceptibilities of the whole material will then be deduced from those of the individual oscillator types through the Lorentz-Lorenz model. Figure \ref{n_i_spectral} displays the result of this treatment in the case of air.

\section{Conclusion}
In conclusion, we have explicitly derived a generalization of the common form of Sellmeier equations, in both frequency-degenerate and non-degenerate systems. This generalization provides the spectral dependence of the non-linear susceptibility of any order, which is fully defined by the knowledge of the linear dispersion of the medium. As a consequence, the Miller formula (\ref{Miller}) \cite{Miller64} can be generalized to any order of non-linearity and any tensor element of non-linear susceptibilities as soon as the material has at least K$\times$C$_2$ symmetry and negligible absorption. In particular, the spectral dependence of non-linear refractive indices of any order can be obtained from their value at one single frequency and the dispersion curve of the medium. Such knowledge is of particular value in nonlinear optics implying confined light leading to very-high intensities, as, \emph{e.g.} in fiber optics \cite{DudleyGC06} filamentation \cite{ChinHLLTABKKS05, BergeSNKW07, CouaironM07, KasparianW08}, or photonic crystals \cite{Soljacic04}.

\section*{Acknowledgments}
This work was supported by the Swiss NSF (contracts 200021-116198 and 200021-125315).

\bibliographystyle{unsrt}

\end{document}